\documentclass[pra,reprint]{revtex4-1}

\newcommand{\ket}[1]{\left\vert#1\right\rangle}
\newcommand{\bra}[1]{\left\langle#1\right\vert}

\newcommand{\expect}[1]{\left\langle#1\right\rangle}

\usepackage{graphicx}
\usepackage{epstopdf}
\usepackage{setspace}

\begin{document}

\title[]{Direct measurement of the system-environment coupling as a tool for understanding decoherence and dynamical decoupling}

\author{Ido Almog$^1$}
\author{Yoav Sagi$^1$}
\author{Goren Gordon$^2$}
\author{Guy Bensky$^2$}
\author{Gershon Kurizki$^2$}
\author{Nir Davidson$^1$}

\affiliation{$^1$Department of Physics of Complex Systems, Weizmann Institute of Science, Rehovot 76100, Israel}
\affiliation{$^2$Department of Chemical Physics, Weizmann Institute of Science, Rehovot 76100, Israel}

\begin{abstract}
  Decoherence is a major obstacle to any practical implementation of
  quantum information processing. One of the leading strategies to
  reduce decoherence is dynamical decoupling --- the use of an
  external field to average out the effect of the environment. The
  decoherence rate under any control field can be calculated if the
  spectrum of the coupling to the environment is known. We present a
  direct measurement of the bath coupling spectrum in an ensemble of
  optically trapped ultracold atoms, by applying a spectrally
  narrow-band control field. The measured spectrum follows a
  Lorentzian shape at low frequencies, but exhibits non-monotonic
  features at higher frequencies due to the oscillatory motion of the
  atoms in the trap. These features agree with our analytical models
  and numerical Monte-Carlo simulations of the collisional bath. From
  the inferred bath-coupling spectrum, we predict the performance of
  well-known dynamical decoupling sequences: CPMG, UDD and CDD. We then
  apply these sequences in experiment and compare the results to predictions,
  finding good agreement in the weak-coupling limit. Thus, our work establishes experimentally the
  validity of the overlap integral formalism, and is an important step
  towards the implementation of an optimal dynamical decoupling
  sequence for a given measured bath spectrum.
\end{abstract}

%\submitto{\JPB}

\maketitle

\section{Introduction}
In any implementation of quantum information processing by qubits
\cite{QCQI} it is crucial to keep the qubits coherent for long periods
of times. The qubits, however, are never completely isolated from the
environment. This coupling to the environment means that after some
time an entanglement between the system (the qubits) and the
environment (all other degrees of freedom) is established. Usually the
environment (bath) consists of many degrees of freedom which are not
controlled. This means that one has to trace over these degrees of
freedom to obtain the state of the system.  The entanglement to the
bath and the tracing leave the system in a separable classical state.
Obviously, the resulting loss of coherence (decoherence) is one of the
major obstacles towards the successful implementation of quantum
information processing.

Over the years, several strategies have been suggested to cope with
the decoherence problem. Clearly, the first thing to do is to minimize
the coupling to the environment. Borrowing ideas from classical error
correction theory, it was shown that by encoding a logical qubit in
several physical qubits, it is possible to correct for errors which
are introduced in the calculation process \cite{QCQI}. Quantum error
correction protocols can correct errors up to some maximal rate. The
upper bound on the error probability of a quantum gate depends on the
details of the error correction protocol, but the typical values range
from $10^{-4}$ to $10^{-2}$. The gate can also be an identity gate
which describes the storage of information in a quantum memory. Such a
memory is essential in the architecture of a quantum network in order
to enable scalability \cite{Duan2001}.

Dynamical decoupling (DD) is a technique that was developed to further
reduce the error rate below the fault tolerant threshold
\cite{Viola1998,PhysRevLett.82.2417,PhysRevLett.85.2272,PhysRevLett.95.180501,uhrig2007}.
The main idea of DD is to use external control fields that induce
rotations of the qubit in the Bloch sphere such that the overall
decoherence is reduced.
This method was first considered in the context of nuclear magnetic
resonance \cite{Kubo1961,JPSJ.9.935,Haeberlen_NMR_book_1976}. In the
field of quantum information processing, the pursuit for fault
tolerance has pushed forward the development of a similar formalism
for controlling the decoherence of noisy qubits
\cite{Viola1998,PhysRevLett.82.2417,PhysRevLett.95.180501}. The
simplest example of DD is the well-known Hahn echo technique: a single
population inverting pulse ($\pi$-pulse) is introduced exactly at half
the final observation time \cite{Hahn1950}. The echo technique is
widely used and very efficient in counteracting the effect of a
quasi-static coupling to the environment. Once the coupling spectrum
extends to higher frequencies, the echo technique fails and more
elaborate control sequences are needed.

A strategy aimed at maximizing the decoupling has been developed by Kofman and Kurizki \cite{Kofman2001,PhysRevLett.93.130406,PhysRevA.76.042310,gordon2007,PhysRevLett.97.110503,PhysRevLett.104.040401}, based on a formula that relates the decoherence rate to the overlap of the bath coupling
spectrum on the control power spectrum. This overlap can be minimized
under the constraint of the available control field energy \cite{gordon2007,PhysRevLett.104.040401}. In
order to successfully implement this decoupling control strategy,
detailed knowledge of the bath-coupling spectrum is required. In this
work, we explain how to measure this spectrum, and then use it to
calculate the coherence of the system after a DD sequence has been
applied. The experiments we present are performed in an ensemble of
ultra-cold atoms held together by a potential induced by a far-off
resonance laser field. This system is used not only for demonstration
purposes, but also because it can serve as a quantum memory \cite{Kuzmich2003,Chou2005,Yuan2008,Zhao2009,schnorrberger:033003,zhang:233602}.

Two of the most widely used physical qubits are photons and neutral
atoms.  Photons are easy to produce, manipulate and transport. They
interact weakly with their environment and therefore can remain
coherent for long travel distances. This last advantage is also their
disadvantage: interactions between photons are usually very small,
making the implementation of an all-optical two-qubit gate very
difficult. Atoms, on the other hand, are easy to keep in one place,
and can interact strongly with other atoms and electromagnetic fields.
It is therefore sensible to use atoms as ``stationary qubits'' for
storage and manipulation and use photons as ``flying qubits'' that
carry the information between distant sites.

One of the controlled schemes of interaction between atoms and photons
is electromagnetically induced transparency (EIT) \cite{Lukin2003}. In
this scheme, atoms with a lambda-shape energy structure interact with
two light fields called ``pump'' and ``probe''. The pump is usually
much stronger than the probe, and is used to control the interaction
strength between the probe and the atomic ensemble. Turning off the
pump while the probe is propagating in the atomic ensemble leads to a
conversion of the photonic excitation into the coherence between the
two low-lying states of the atoms. This is sometimes called ``storage
of light'', although only the coherence which was carried by the light
is actually stored in the ensemble. The beauty of this conversion
process is that it is reversible, which makes the atomic ensemble a
true memory.

In order to increase the efficiency of the storage and retrieval
processes, it is desirable to work with atomic ensembles with high
optical depth \cite{PhysRevLett.82.4611,gorshkov2007}. This is because
the coupling of the atoms to the external electromagnetic field scales
as the square root of the number of atoms in a volume where the light
intensity is approximately uniform. Working at high optical depth,
however, usually implies that the atomic density is high and the rate
of inter-particle collisions is large compared to the storage time.
The coherence properties of the atomic ensemble are markedly changed
due to elastic collisions \cite{PhysRevLett.105.093001}. From the
point of view of a particular atom, other atoms can be regarded as the
bath. The collisions with other atoms define the nature of the
coupling to the bath. The three important physical quantities in the
description of the collisional bath are: the collision rate, the
inhomogeneous dephasing rate (i.e. the dephasing without collisions)
and the harmonic confinement oscillation frequency. The spectral
behavior of the collisional reservoir stems from the interplay between
these quantities.

The goal of this work is to present a general method of measuring the
bath-coupling spectrum, and demonstrate its usefulness in calculating
the result of any control field. We start by reviewing the mathematical formalism
of the  overlap integral spectrum and the power spectrum of the control sequence that acts as
filter function (Section~\ref{section_fidelity}). Using this formalism
it is possible to calculate from the bath-coupling spectrum the
coherence at some observation time with any control sequence, as long
as the so-called 'weak coupling' limit applies. In particular, it can
be used with a null control sequence, which simply describes a
dephasing process. For a known filter function of the control field,
it is possible to de-convolve the overlap integral to obtain the
bath-coupling spectrum. We present this technique for a constant-power
continuous control field in
Section~\ref{section_bath_measurement_technique}, and illustrate it by
full 3D Monte-Carlo simulations. We then employ the technique to
measure directly the collisional bath spectrum in our cold atomic
ensemble (Section~\ref{section_bath_measurement}).  The measured bath
shows interesting non-monotonic behavior due to the oscillatory motion
of the atoms in the trap. Once the bath-coupling spectrum is known, it
is interesting to test its implications for the outcome of other DD
sequences. This is done in Section~\ref{section_dynamical_decoupling},
where we use the measured bath and the overlap-integral formalism to
calculate the coherence under the well-known CPMG, UDD and CDD
sequences. We then experimentally apply these DD sequences and compare
the results with the predictions of the theoretical calculations based
on the measured bath. We conclude and give our outlook in
Section~\ref{section_conclusion}.

\section{The spectral overlap integral approach to the fidelity
  calculation}\label{section_fidelity}

In this section we review the formalism \cite{Kofman2001,PhysRevLett.93.130406,PhysRevA.76.042310,gordon2007,PhysRevLett.97.110503,PhysRevLett.104.040401} which is useful to calculate the ensemble
coherence at a given time. Here we give a simplified version of this
approach assuming an effective two-level stochastic Hamiltonian weakly
coupled to the environment.

We consider the same model as was described in Ref. \cite{our_process_tomography}. Our ensemble consists of two-level systems (TLS), and due to inhomogeneities in the external environment, the transition energy of each TLS is different than the frequency it may have in free space (we shall call this difference the detuning). We reduce the full many-body Hamiltonian to an effective single particle Hamiltonian given by
\begin{equation}\label{Hamiltonian3}
\hat{H}=\hbar\left[\omega_0+\delta(t)\right]\ket{2}\bra{2}+\hbar\Omega(t)\ket{2}\bra{1}+h.c. \ \ ,
\end{equation}
where $\omega_0$ is the free space transition frequency between the two states, $\delta(t)$ is the detuning, and $\Omega(t)$ is a classical external control field which is going to be used for the DD. The detuning is a random function of time, and we assume it is averaging to 0 over different realizations of the Hamiltonian. The reduced density matrix can be calculated by
\begin{equation}
\rho(t)=\overline{\ket{\phi(t)}\bra{\phi(t)}} \ \ ,
\end{equation}
where $\phi(t)$ is the state of the system at time $t$ with a specific realization of the stochastic Hamiltonian, and the average is taken over many realizations. $\ket{\phi(0)}$ is the initial state of the system, and ultimately our goal is to conserve this state in the ensemble for as long as possible. To characterize how well this goal is achieved it is instructive to introduce the fidelity function:
\begin{equation}
\mathcal{F}=\bra{\phi(0)}\rho(t)\ket{\phi(0)} \ \ ,
\end{equation}
which starts at $1$ and decays to $0.5$ at long times.

To make the model clearer, we now relate these definitions to atomic
ensembles \cite{PhysRevLett.105.093001}. In our system, ultracold
atoms are confined by an external optical potential. Although the
atoms have many energy levels, we confine our attention to two
low-lying states with negligible spontaneous decay.  The potential is
not exactly the same for the two states. The main source inhomogeneous
broadening is that the energy difference between the two sates, which
is proportional to the total energy of the atom, is changing in space
\cite{kuhr:023406}. The phase space density of our ensemble is low
enough to be considered classical (far from quantum degeneracy). The
motion of the atom in the external potential causes the transition
frequency to change in time. Since the ensemble consists of many
atoms, each with a different trajectory, measuring the coherence of
the ensemble gives the averages discussed above in a single shot. In
the experiment, the coherence is measured in a time-domain Ramsey-like
experiment, as described in Ref.  \cite{PhysRevLett.105.093001}.

We shall define two functions which are going to be useful later for
the calculation of the fidelity. First, the bath coupling spectrum
characterizes the spectral content of the coupling of the system to
the bath and is defined as
\begin{equation}\label{bath_function}
G(f)=\int_{-\infty}^{\infty}{e^{-2\pi i f \tau}\expect{\delta(t)\cdot \delta(t+\tau)}d\tau} \ \ ,
\end{equation}
where $\expect{...}$ stands for the averaging over many realizations
of $\delta(t)$. $G(f)$ is the Fourier transform of the time
correlation function of $\delta(t)$. The second function we define is
a 'filter function' which characterizes the power spectrum of the
control sequence:
\begin{equation}
F_t(f)=\left\vert \int_0^t{e^{-2\pi i f t}\cdot \cos\left(\int_0^t{\Omega(\tau)d\tau}\right)dt }\right\vert^2 \ \ .
\end{equation}
Note that although $F_t(f)$ is defined in the frequency domain, it does depend on the observation time $t$.

A simple expression for the fidelity at time $t$ can be obtained under
the \emph{weak coupling} assumption, namely that the fidelity decay
during the bath correlation time is negligible. In this case the
fidelity is given by the overlap integral
\cite{Kofman2001,PhysRevLett.93.130406,gordon2007}:
\begin{equation}\label{eq_for_F_t}
\mathcal{F}(t)=\frac{1}{2}(1+e^{-R(t)t}) \ \ ,
\end{equation}
where $R(t)$ is the decay rate given by the overlap integral
\begin{equation}\label{decay_rate_integral}
R(t)=\frac{2\alpha}{t}\int_0^t{G(f)F_t(f)df} \ \ ,
\end{equation}
where $\alpha$ is a constant between 0 and 1, depending on the
assumptions for the statistics of the initial state, in particular,
$\alpha=\frac{1}{4}$ for an initial state which is an equal
superposition of the two internal states. Generally speaking, $R(t)$
is not constant and therefore the decay in not exponential: a signature of the non-Markov time-domain.

A related quantity is the ensemble coherence, which is defined to be
the normalized off-diagonal element of the reduced density matrix
\cite{cywinski:174509}. In the overlap integral framework, the
coherence is given by $\mathcal{C}(t)=e^{-R(t)t}$. In the Bloch sphere
representation, the coherence is given by the length of the Bloch
vector, which is usually measured by quantum state tomography
\cite{QCQI,our_process_tomography}.

\section{How to measure the bath coupling spectrum ?}\label{section_bath_measurement_technique}

In order for Eqs.~(\ref{eq_for_F_t}-\ref{decay_rate_integral}) to be
useful, one has to know the bath coupling spectrum $G(f)$, preferably
from experimental data. This spectrum can be inferred from
Eq.~(\ref{decay_rate_integral}) by measuring the coherence for a given
pulse sequence with a known $F_t(f)$. Obviously, if $F_t(f)$ where a
Dirac $\delta$ function, the decay rate would be linearly proportional
to the bath coupling spectrum. A good approximation to a $\delta$
function can be obtained if the control field is nearly-continuous and
on-resonant field. This control field causes Rabi oscillations with a
frequency $f_0$ which depends on its strength. The resulting filter
function is a sinc-function centered around $f_0$. To be more precise,
the filter function is given by
\begin{equation}
F_t(f)=\frac{1}{4}t\left[\mathrm{sinc}^2(t(f-f_0))+ \mathrm{sinc}^2(t(f+f_0))\right] \ \ ,
\end{equation}
where $t$ is the pulse duration. For this pulse, assuming $f_0T\gg1$,
Eq.~(\ref{decay_rate_integral}) yields the following decoherence rate:
\begin{equation}\label{decay_for_long_Rabi}
        R(f_0)\cong\frac{1}{4}G(f_0) \ \ .
\end{equation}
We see, then, that by scanning the strength of the control fields
(thus scanning $f_0$) and measuring the decay of coherence, it is
possible to directly measure the bath coupling spectrum.

To illustrate the method and check its sensitivity to the
weak-coupling assumption, we have run a 3D Monte-Carlo simulations of
a cold atomic ensemble trapped in an optical potential. The simulation
solves for the classical Newtonian motion of $3500$ atoms. The initial
conditions of the atoms are drawn from a Boltzmann distribution,
assuming a temperature of $7\mu K$. As explained previously, the
fluctuations in our system originate from elastic collisions between
the atoms, and it is therefore necessary to include them in the
simulation. In order not to run into computational complexity
problems, we have developed a mean-field technique
\cite{remark_collisions}. Once the trajectories of all the atoms are
calculated, we calculate the energy shift of the internal states
induced by the external potential along the trajectory of each atom.
In our $\gamma=1.06\mu m$-wavelength dipole trap, the differential
shift of the internal states is $6.6\cdot10^{-5}$ times the overall
potential \cite{kuhr:023406}. These energy shifts are then used to
solve the Bloch equations \cite{eberly_book} in the presence of the
control field. We calculate the decoherence rate by computing the
length of the Bloch vector at different times and fit it to a decaying
exponent.

\begin{figure}
    \centerline{\includegraphics[width=8cm]{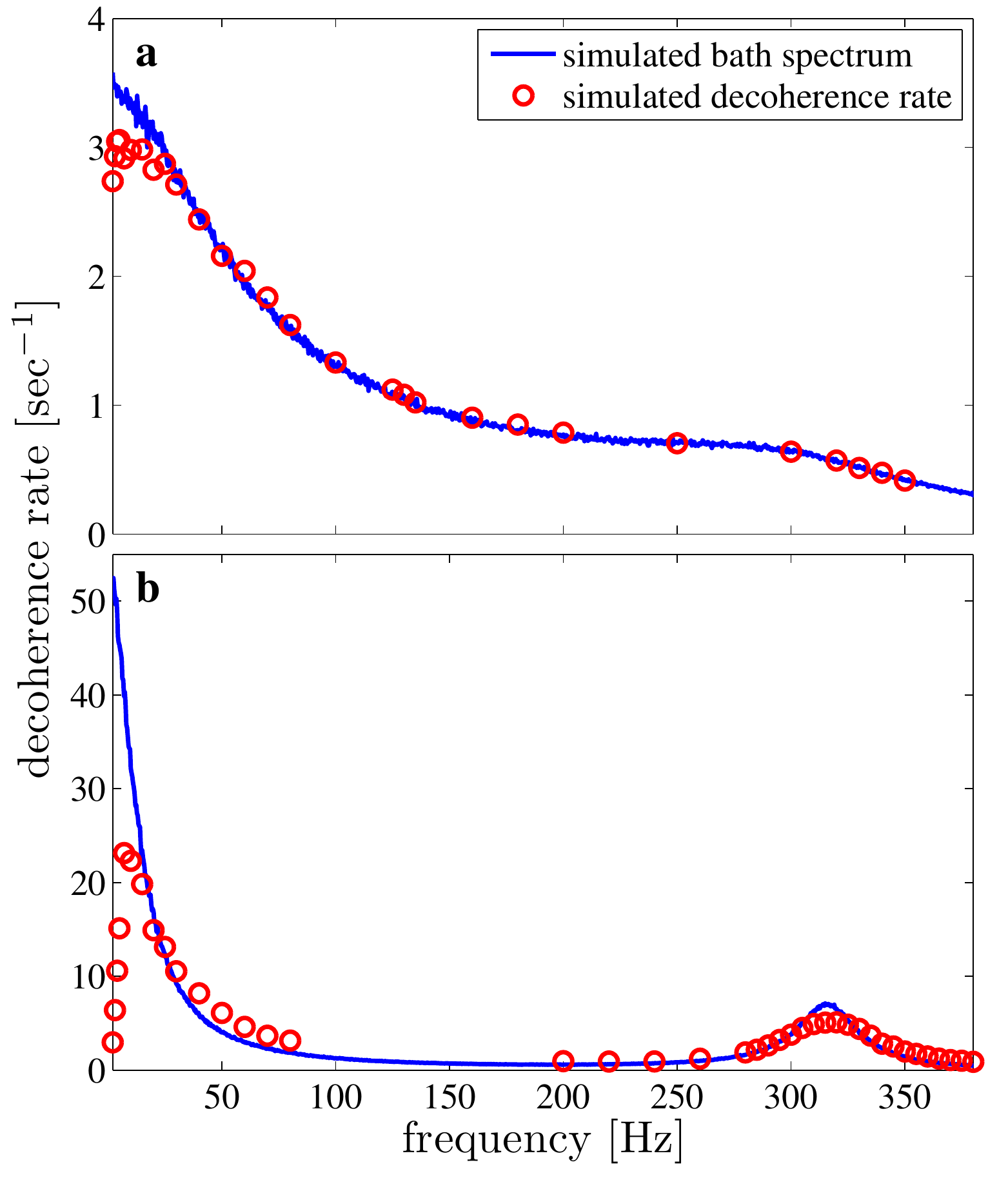}}
    \caption{Numerical simulations of the bath-coupling spectrum inferred from decoherence
      measurements as compared to the direct calculation. We simulate the classical motional of $^{87}Rb$ atoms in the trap. The blue solid line is the bath spectrum as calculated directly from the detunings
      along the atomic trajectories using Eq.~(\ref{bath_function}).
      The red circles are the decoherence rates of atoms subject to a
      continuous field (long pulse) with the Rabi frequency plotted on
      the x-axis. The conditions of the simulations are a temperature
      of $T=7\mu K$, radial and axial oscillation frequencies of
      $2\pi\cdot600$Hz and $2\pi\cdot160$Hz, respectively, and number
      of atoms, $N=2 \cdot 10^6$ for graph \textbf{(a)} and
      $N=1.5\cdot 10^5$ for graph \textbf{(b)}.  The deviation of low
      frequencies of the simulated decoherence measurements from the
      directly simulated spectrum is due to the breakdown of the
      weak-coupling assumption. The number of atoms is different in
      the two graphs and so is the frequency at which the breakdown
      occurs.  In graph \textbf{(a)} the non-monotonic spectral
      feature (peak) is smeared due to the much larger Lorentzian
      width.}
    \label{decay_spectrum_relations_sim}
\end{figure}

We simulate a measurement of the bath spectral function as described
above by a continuous control field, in two parameter regimes which
correspond to weak and strong coupling. The results of the simulations
are depicted in Figure~\ref{decay_spectrum_relations_sim}. At low
frequencies, the spectrum follows a Lorentzian curve which is expected
since the fluctuations originate from a scattering process with
Poisson statistics \cite{our_process_tomography}. Also apparent in the
spectrum is a non-monotonic feature, due to the oscillatory motion of
the atoms in the trap (for more details see the next section). The
only difference between the two simulations is the number of atoms,
resulting in different correlation times of the bath.  It is
interesting to note that by increasing the number of atoms one also
decreases the decoherence rate at low frequencies, a result of
collisional narrowing (see Ref. \cite{PhysRevLett.105.093001} for more
details).  Thus, increasing the number of atoms increases both the
collision rate and the coherence time, thereby pushing the ensemble
deeper into the weak-coupling regime. Note that the relevant
decoherence rate for considering whether the system in weakly coupled
is the decoherence rate obtained {\em with} the decoupling pulse. This
means that a system can be in the strong-coupling regime at low
frequencies and in the weak-coupling regime at high frequencies,
whence deviations from the calculated spectrum are larger at low
frequencies.

\section{Measurement of the bath coupling spectrum}\label{section_bath_measurement}

We have used the technique described in the previous section to
measure the bath coupling spectrum in an ensemble of colliding
ultra-cold atoms. In the experiment, $^{78}Rb$ atoms are trapped in an
optical potential created by a far-off-resonance laser with a
wavelength of $1.064\mu m$. Initially $\sim10^9$ atoms are trapped and
cooled in a magneto-optical trap, and further cooled by the Sisyphus
\cite{limits_sisyphus_1991} and Raman-sideband techniques
\cite{Kerman2000}. We then use rapid adiabatic passage with a constant
RF radiation and a ramped-up magnetic field to transfer the atoms from
the state $\ket{5^2S_{1/2},F=1;m_f=1}$ to the state $\ket{5^2S_{1/2}
  F=1 m_F=-1}$for $\sim 69\%$ of the atoms, and the rest to the state
\mbox{$\ket{5^2S_{1/2},F=1;m_f=0}$}. The thermodynamic parameters of
the ensemble are measured by absorption-imaging of the cloud.

The optical trap consists of two beams crossing at an angle of
$28^\circ$ after passing through a zoom system capable of controlling
their waist. We start by collecting the atoms in a $180\mu m$-waist
beam, and then dynamically compress the waist down to $50\mu m$. The
polarization of the two crossing beams is parallel to the magnetic
field, and their frequency differs by $120$MHz, in order to eliminate
standing waves.  In this experiment we choose to work with the two
states $\ket{1}=\ket{5^2S_{1/2} F=1 m_F=-1}$ and
$\ket{2}=\ket{5^2S_{1/2} F=2 m_F=+1}$. At the applied magnetic field
of $3.23$ Gauss these states are Zeeman insensitive to magnetic field
fluctuations to the first order \cite{Harber2002}. The two states are
separated by an energy of $2\pi\hbar\cdot 6.833$GHz, but are slightly
affected by the differential AC Stark shift of the dipole trap
\cite{kuhr:023406}.  As explained above, for a moving atom this shift
is time-dependent and follows the trajectory.

Since $\Delta m=\pm 2$ between the two internal states, the external
control $\Omega(t)$ has to be effected by a two-photon transition. We
employ RF radiation at $2.15$MHz and microwave radiation at
$6.832527928$GHz, which is chosen such that both frequencies are
detuned by $90kHz$ from the intermediate level $\ket{5^2S_{1/2} F=2
  m_F=0}$. The maximum Rabi frequency we achieve is $\Omega=2\pi\cdot
1000$Hz. To detect the state of the atoms we use a state-selective
fluorescence-detection scheme, similar to the one described in Ref.
\cite{Khaykovich2000}.

Two main difficulties arise from driving the system by a strong
control field. First, at high Rabi frequencies any inaccuracies from
shot to shot in the field strength translate into noise in the atomic
population. For example, an average noise of $1\%$ in the Rabi
frequency transforms into $\sim100\%$ noise in the atomic population,
if we drive our system at $\Omega=2\pi\cdot 100$Hz for 1sec.  Strictly
speaking, this noise in the control field results in a reduction of
the fidelity. In order to avoid this difficulty, we randomize
completely the initial population difference by an additional pulse
that rotates the atomic state by a random uniformly distributed angle
and analyze $\sim30$ data points taken for the same pulse power using
envelope spectroscopy. The corresponding Bloch-vector length is
estimated using the maximum likelihood estimator. Since the angle of
the Bloch vector is now completely random we can assume that its
z-component is $C\sin(\Phi)$, where $\Phi$ is a uniformly distributed
random phase. The results of the maximum-likelihood algorithm that
extracts the Bloch vector length C coincide with the extreme values of
the measured z-component (the population inversion).

\begin{figure}
\centerline{\includegraphics[width=8cm]{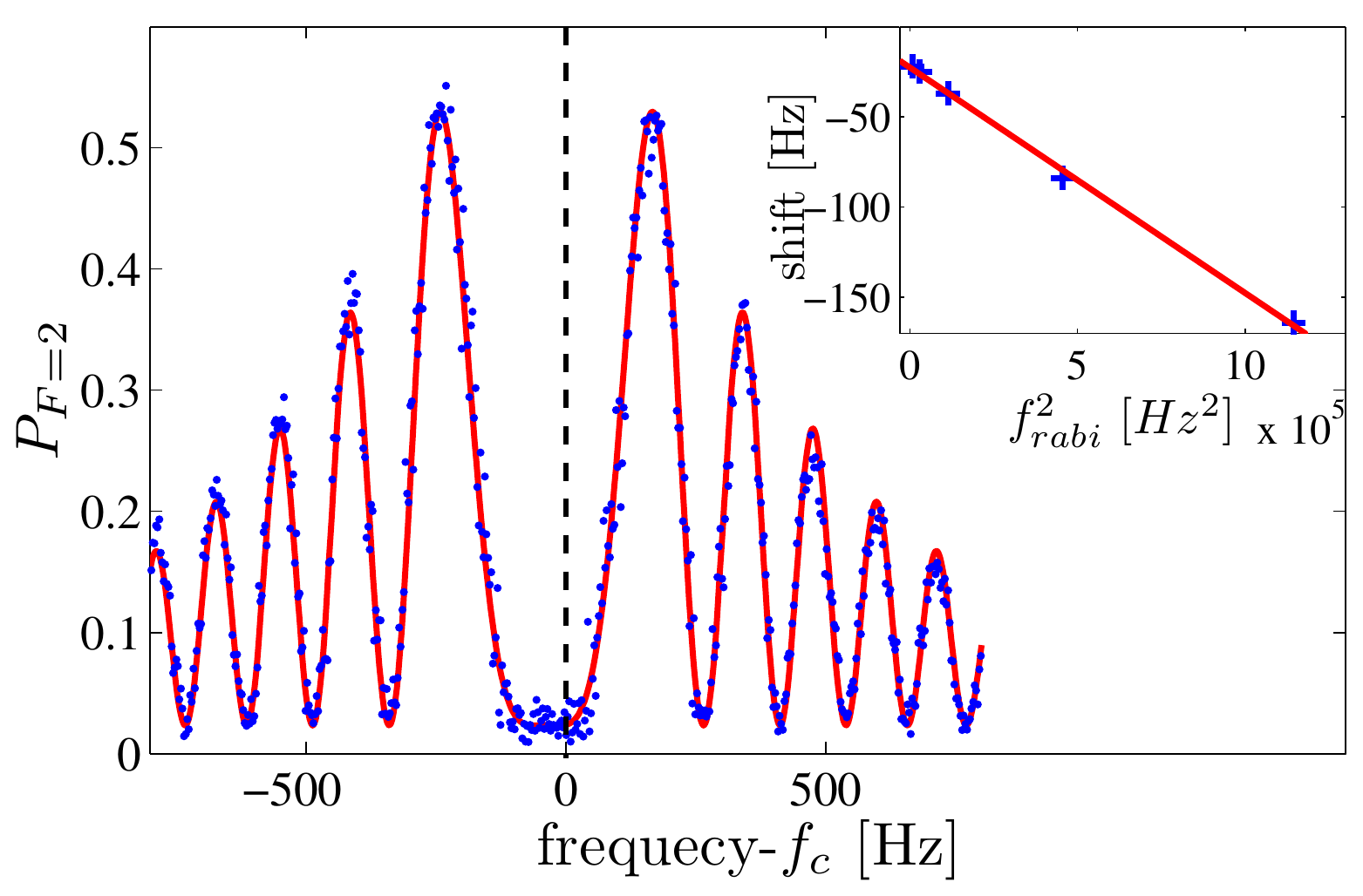}}
\caption{Calibration of the microwave dressing effect. The y-axis
  shows the population at F=2 as a function of the control pulse
  frequency shifted by $f_c$, the center frequency without the
  dressing. The data was taken with a Rabi frequency of
  $\Omega=340$Hz. We fit the data to the theoretical curve
  $A\frac{f^2}{(f-f_0)^2+f^2}[1+\cos(\phi+T\sqrt{(f-f_0)^2+f^2})]$,
  where $T$ is the duration of the Rabi pulse and $f$ is the field
  frequency. The fit gives $f_0$, the dressing for the applied Rabi
  frequency. In this example, the shift is $-37$Hz. The inset shows
  the microwave dressing shift as a function of the Rabi frequency
  squared.}
\label{long_rabi_spec}
\end{figure}

The second difficulty is due to the energy-level shift induced by the
control field \cite{PhysRevLett.24.861}. This ``field dressing'' of
the levels depends on the microwave field strength (we do not observe
the shift due to the rf field) . When the bath spectrum is measured
using a continuous control field, this effect causes the control to
have a detuning which depends on its strength (and therefore on its
Rabi frequency). We measure the shift by applying a long pulse which
induces Rabi oscillations and detect the population at state $\ket{2}$
while scanning the frequency of the microwave field. Such a
measurement is shown in Figure~\ref{long_rabi_spec}. This data yields
the relative-level shift for a given Rabi frequency. We then repeat
this measurement for several Rabi frequencies and obtain a calibration
graph, shown in the inset of Figure~\ref{long_rabi_spec}.  As expected
from theory, the dressing at small field strengths is linearly
proportional to the control-field Rabi-frequency squared
\cite{PhysRevLett.24.861}.

\begin{figure}
  \centerline{\includegraphics[width=8cm]{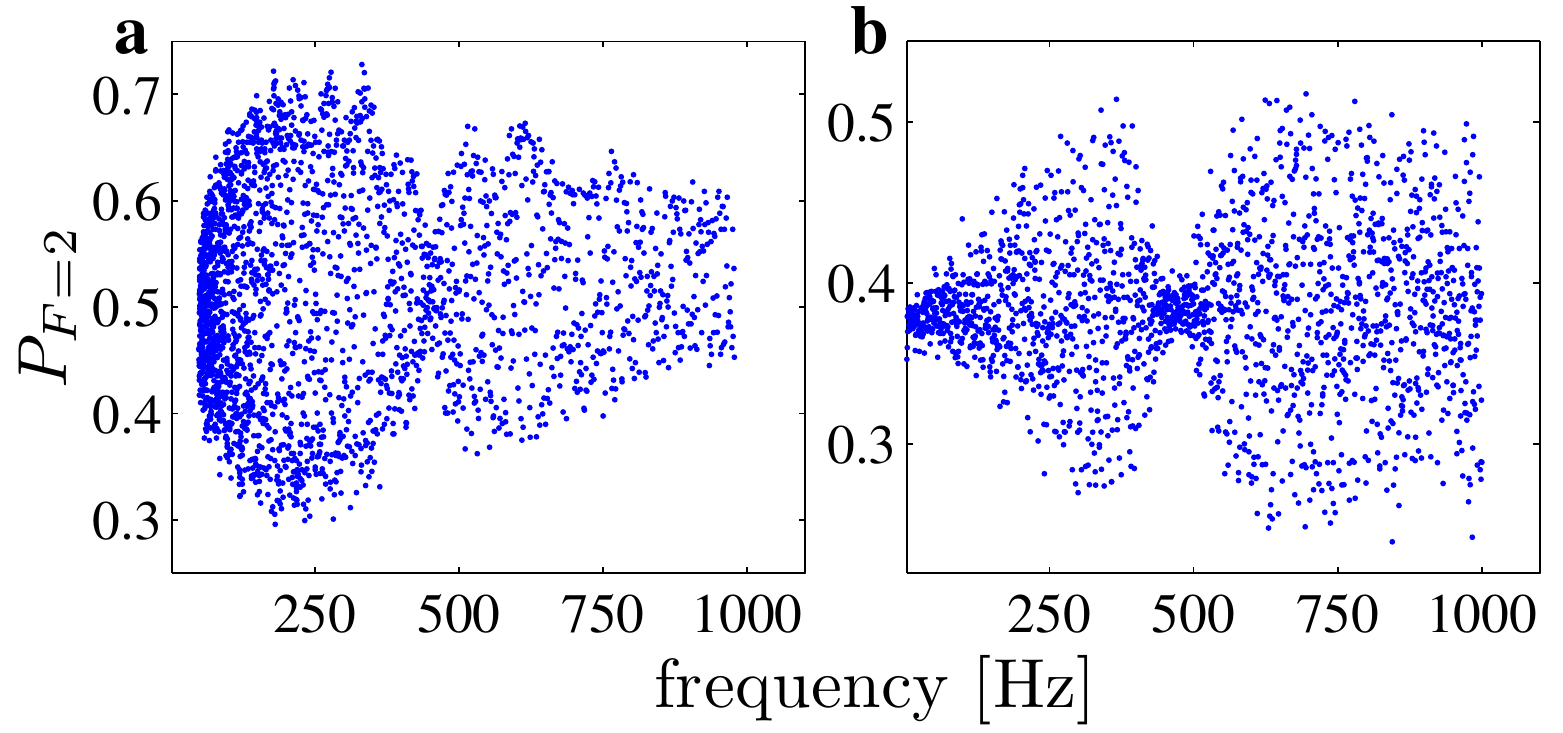}}
  \caption{Raw data of coupling spectra. The points are the population
    at F=2 as a function of the control pulse Rabi frequency. The
    experiment was done for $160$k atoms in a trap with a radial and
    axial oscillation frequencies of $2\pi\cdot 910$Hz and $2\pi\cdot
    240$Hz, respectively. The field dressing results in the reduction
    of the contrast which is dominant in high frequency.  On the right
    we show that after correction for this field-dressing the contrast
    does not decay at higher frequencies even after $1$sec. The
    contrast between the two graphs is different since the observation
    time is $0.4$sec for the left graph and $1$sec for the right
    graph. The difference in the mean population stems from m-changing
    transitions in the $F=2$ hyperfine level. The coherence is
    calculated by a maximal likelihood method (see text), and is
    determined by the envelope of the scattered data points. }
  \label{fish_wandwo_mwdressing}
\end{figure}

In Figure~\ref{fish_wandwo_mwdressing} two examples of raw data from a
bath-spectrum measurement are shown. The scattering of the data points
is a consequence of our randomization technique explained above.
The bath-coupling spectrum at a given frequency is
related to the envelope of the scattered points. The data presented in
Figure~\ref{fish_wandwo_mwdressing}\textbf{a} was taken without
changing the control field frequency (not to be confused with its
strength which controls the Rabi frequency).

The second difficulty described above leads to a substantial reduction
of the contrast at higher Rabi frequencies due to the detuning of the
control field from resonance, caused by the strong field-dressing of
the levels. In order to eliminate this effect, we change the frequency
of the microwave field as its power changes to compensate for the
field dressing. The result is shown in
Figure~\ref{fish_wandwo_mwdressing}\textbf{b} to restore contrast at
high frequencies. In general, this correction should also be applied
in other DD sequences. In the cases we present below, however, the
dressing effect is very small and unimportant.

\begin{figure}
\centerline{\includegraphics[width=8cm]{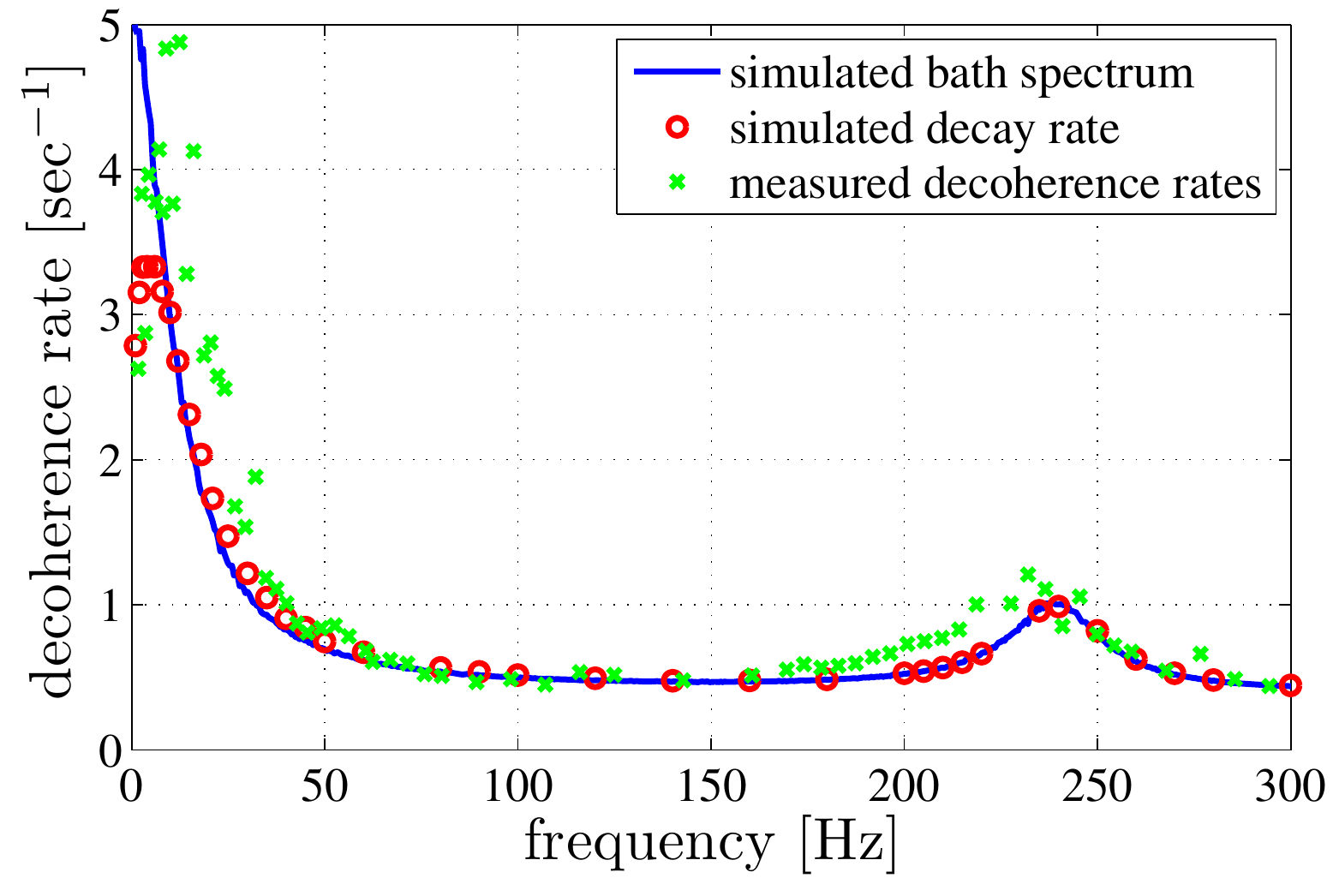}}
\caption{Comparison of the measured bath spectrum (plus signs) to the
  simulated spectrum (solid line) and decoherence rate (circles). The
  conditions for the experiment and simulation are $340k$ atoms at a
  temperature of $3.5\mu K$, and with the power of the trapping laser
  being $1.7$W, creating a trap with a radial and axial oscillation
  frequencies of $2\pi\cdot 450$Hz and $2\pi\cdot 120$Hz,
  respectively. The simulated spectrum was shifted by a constant of
  $0.4s^{-1}$ to better fit the data. This shift accounts for $T_1$
  processes and a bias which arises due to noise in the measurements.}
\label{simulation_vs_exp}
\end{figure}

To obtain the bath-coupling spectrum, we repeat the measurement
described above with at least three different Rabi pulse durations.
From each such experiment we get the coherence at that time for all
Rabi frequencies. For each Rabi frequency we fit the decay of the
coherence at different times to a decaying exponent, from which we
obtain the decoherence rate. As explained above, the fitting to an
exponent is justified in the weak-coupling regime. In
Figure~\ref{simulation_vs_exp} we plot the experimentally measured
decoherence rate as a function of frequency. Similarly to the
simulations presented in Figure~\ref{decay_spectrum_relations_sim},
the measured spectrum follows a Lorentzian at low frequencies and
shows a non-monotonic feature at twice the axial oscillation
frequency. This feature arises due to the rapid oscillation of the
atoms in the potential. The reason is that the detuning function
$\delta(t)$ always has a component at twice the oscillation frequency
of the trap (since the detuning is proportional to the potential which
scales as $x^2$). Measuring at twice this frequency couples to this
component and produces higher decoherence \cite{2011arXiv1101.4885K}.

For comparison we plot in Figure~\ref{simulation_vs_exp} the results
of our Monte-Carlo simulation with the same parameters as those
measured in the experiment. The agreement between the measured and
simulated spectrum is quite satisfactory. We observe some shift
between the two spectra in the position of the non-monotonic feature.
This is probably due to small inaccuracies in the measured laser power
and waist used in the simulation.  The measured width of the second
peak is larger in the experiment, probably due to the anharmonicity of
the Gaussian confining potential which is not simulated.

Figures~\ref{decay_spectrum_relations_sim}-\ref{simulation_vs_exp}
prove the ability of this technique to
observe small dynamical effects through the spectrum. We note that
although other methods to measure the oscillation frequency of the
trapping potential exist, the use of the spectrum to infer this
quantity is unique in that it does require excitations of the atoms.

\section{Dynamical decoupling and the bath spectrum}\label{section_dynamical_decoupling}

As explained in Section~\ref{section_fidelity}, the bath coupling
spectrum is most useful for calculating the outcome of any DD
sequence. In general, for a given bath spectrum and a set of
constraints (such as the maximum duration, energy or action of the
control pulse), it is possible to devise an optimal decoupling
sequence \cite{gordon:010403,PhysRevLett.104.040401}. There are, however, well-known sequences which
are suitable for certain classes of noise spectra. Here we concentrate
on three such sequences: the Carr-–Purcell-–Meiboom-–Gill sequence
(CPMG) \cite{PhysRev.94.630,meiboom:688,RevModPhys.76.1037}, the
concatenated DD sequence (CDD) \cite{PhysRevLett.95.180501}, and the
sequence developed by Uhrig (UDD) \cite{uhrig2007}. Although in this
work we restrict ourselves to sequences of brief $\pi$-pulses, this is
not required by the theory. To compare the performance of the three sequences
we consider the decoherence they yield for a given number of $\pi$
pulses.

\begin{figure}
\centerline{\includegraphics[width=8cm]{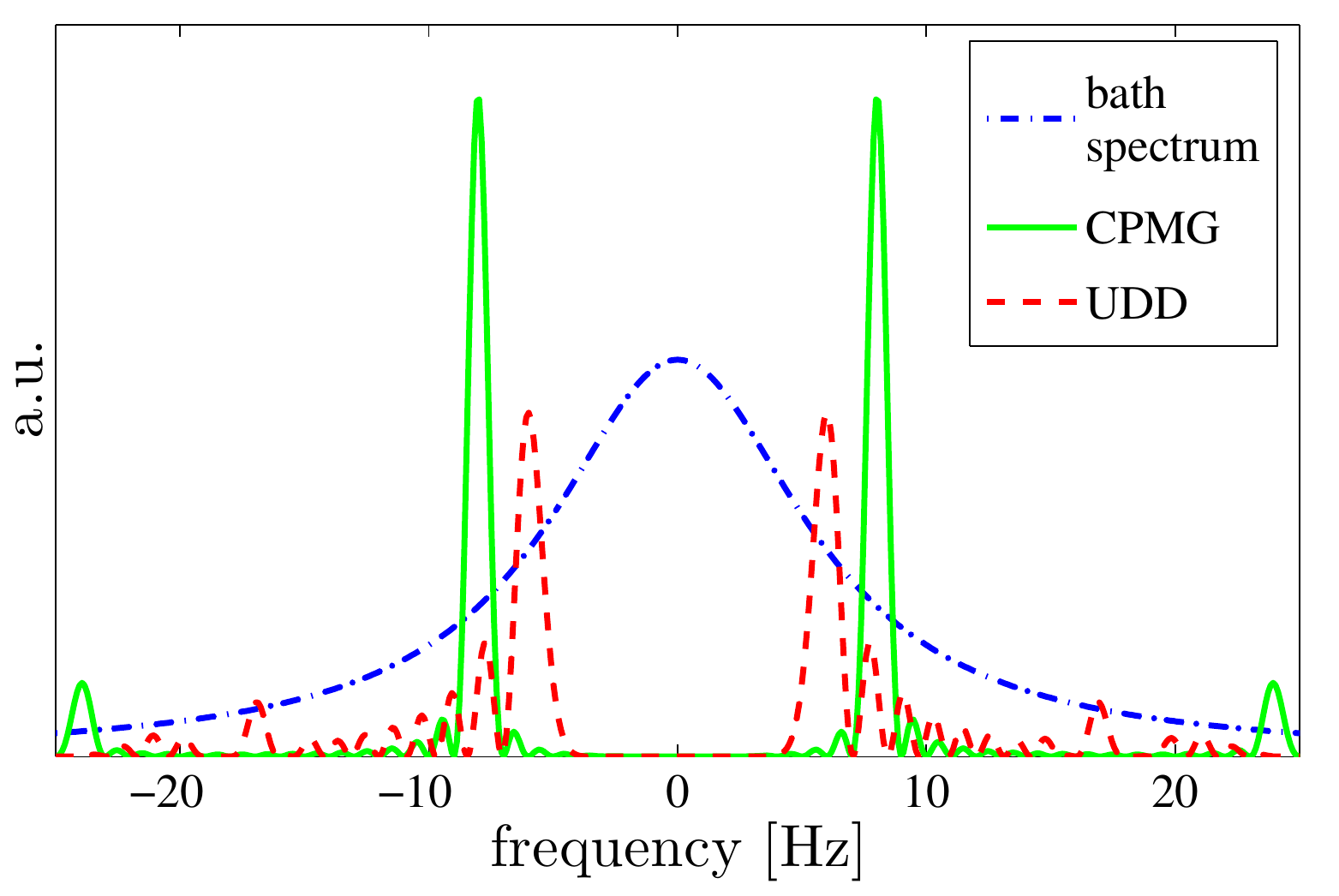}}
\caption{The collisional bath-coupling spectrum, the filter functions
  and their overlap, calculated from the formula
  (Eq.~(\ref{decay_rate_integral})) for two DD
  sequences: CPMG and UDD. The dash-dotted line is a Lorentzian
  spectrum which describes the Poisson statistics of elastic
  collisions. The filter function for the CPMG decoupling sequence
  (solid line) and UDD (dotted line), both consisting of 16
  $\pi$-pulses for an observation time of $1$sec. The UDD spectral
  overlap is larger at lower frequencies compared to the that of CPMG.
  Since the bath-coupling spectrum at these frequencies is higher
  (which means a faster decay), the resulting decoherence rate for UDD
  is larger than for CPMG.}
\label{compare_schematic}
\end{figure}

In Figure~\ref{compare_schematic} we depict schematically the
Lorentzian spectrum of our collisional bath and the filter function
(Eqs.~(\ref{decay_rate_integral})-(\ref{decay_for_long_Rabi})) of the CPMG and UDD sequences with 16 $\pi$ pulses.
The CPMG spectrum has a sharp peak at $f_0=8$Hz, half the pulse rate,
and smaller peaks at lower and higher frequencies whose envelope is
proportional to $1/(f-f_0)^{2}$. There are also peaks at harmonies
centered around $3f_0$ but their area is 9-times smaller than that of the main peak. UDD has
exponentially reduced coupling at low frequencies.  This can be very
useful for bath spectra which have a cutoff at high frequencies
\cite{Biercuk2009}, which is not the case of the Lorentzian spectrum.
Namely, the UDD exponential suppression of the lower sidebands, comes
with a price: for the same number of pulses the maximum coupling is at
lower frequency, hence for many physical coupling spectra, including
Lorentzian spectra, it will sample portions of the spectrum with
larger decoherence. CDD (not shown in the graph) has similar
properties to UDD. It has an exponentially reduced coupling in part of
the spectrum, but its overlap with a Lorentzian is overall the same.

Experimentally, we apply the DD sequences for a duration of $0.4$sec,
on an ensemble prepared with the same conditions as in the previous
section. The atoms are initialized to an equal superposition
$\ket{\Psi}=\ket{1}+\ket{2}$, which we want to preserve. The duration
of a $\pi$-pulse was chosen in these experiments to be $2.3$msec,
corresponding to a maximal Rabi frequency of $217$Hz. To reduce the
noise introduced by the inaccuracies in the control field, we switch
the phase of the control by a $\pi$-phase every two pulses
$\pi,\pi,-\pi, -\pi,...$ \cite{our_process_tomography}. The final
atomic state at the end of the sequence is determined using state
tomography. Our measurement of the population gives only the
z-component of the Bloch vector. In order to measure the other two
components we apply before the measurement an additional $\pi/2$
rotation around the $x$ or $y$ axes. We tune the duration of our $\pi$
pulses in the DD sequences such that the there is only a small
z-component after the sequence. The x and y components are measured by
a Ramsey technique, in which we vary the phase of a final $\pi/2$
pulse and measure the contrast of the fringe. In other words, the
state tomography gives us the ability to tune correctly the $\pi$
pulse duration such that the final state always resides in the
equatorial plane of the Bloch sphere. It is then more accurate to
measure the coherence by applying a final $\pi/2$ pulse with a
controlled phase and detect the population in the z-axis. Finally, we
calculate the coherence time by fitting the decay of the fringe
contrast at different times to a decaying exponent.

\begin{figure}
\centerline{\includegraphics[width=8cm]{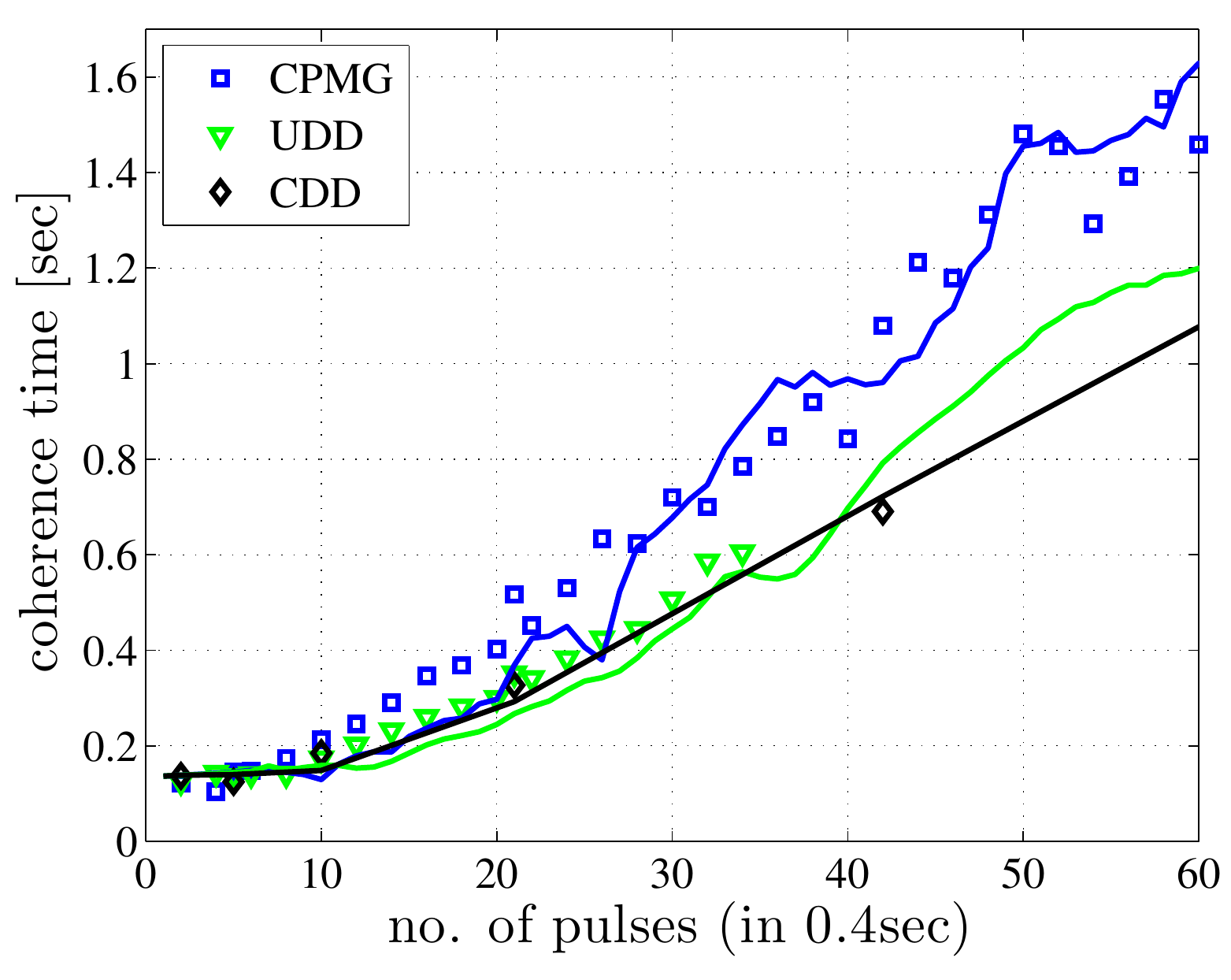}}
\caption{Coherence times under different DD pulse sequences. We
  compare the sequences CPMG (blue squares), UDD (green triangles) and
  CDD (black diamonds) for a given number of pulses at an observation
  time of $0.4$sec. The performance of both UDD and CDD is worse than
  the CPMG, as expected from their overlap with the bath spectrum (see
  Figure~\ref{compare_schematic}). The solid lines is the coherence
  time as calculated based on the overlap integral of the sequences
  filter functions with the measured bath spectrum. The conditions of
  the ensemble are the same as for the measured bath spectrum in the
  previous section. Note that the CDD sequence cannot have any number
  of pulses, and this is the reason why there are only 5 data points
  for this sequence.}
\label{compare_real}
\end{figure}

The results of these measurements are presented in
Figure~\ref{compare_real}. For all DD sequences the coherence time
increases as the number of pulses grows. As expected from the overlap
between the filter functions of the DD sequences and the bath spectrum
(see Figure~\ref{compare_schematic}), the CPMG surpasses both UDD and
CDD, which in turn behave quite similarly. We stress that our ranking
between the DD sequences is only valid for our experimental
collisional bath. Even more interesting than the relative performance
of the DD sequences is to compare them with the theory presented in
Section~\ref{section_fidelity}. To this end, we take the measured bath
spectrum from the previous section (Figure~\ref{simulation_vs_exp})
and use it in the overlap integral formalism to calculate the
predicted outcome of each sequence. There is, however, a depopulation
$T_1$ process in our system which results from m-changing transitions
in the F=2 hyperfine level. We measure this timescale directly to be
$T_1=2.2$sec in our experimental conditions. Since the $T_1$ and $T_2$
processes are not correlated, the total decay is the sum of the decay
rates of both processes. In practice, we subtract from the measured
bath spectrum the bias which results from the $T_1$ process and the
noise in the measurement. By doing so we can calculate only the
decoherence resulting due to the fluctuations of $\delta(t)$ (the
$T_2$ process decoherence rate). We then add to it the decay due to
the $T_1$ process, which is given by $2/T_1$. The total decoherence
rate is depicted as a solid line in Figure~\ref{compare_real}, and
agrees very well with the coherence times that were measured directly.
We attribute the small deviations at low frequencies to a departure
from the weak-coupling regime.

\section{Conclusions and outlook}\label{section_conclusion}

To summarize, we have presented a method of measuring directly the
bath-coupling spectrum, and used it for optically trapped ultracold
atoms. The measured spectrum follows a Lorentzian shape at low
frequencies, which is expected, since the source of fluctuations is
s-wave scattering between the atoms \cite{our_process_tomography}.  At
higher frequencies, though, the spectrum exhibits non-monotonic
features which arise due to the rapid oscillatory motion of the atoms
in the trap. The usefulness of the concept of the bath-coupling
spectrum is tested by its ability to predict the outcome of any DD
control sequence. We have measured directly the performance of three
well-known sequences: CPMG, UDD and CDD. The comparison of these
measurements to a calculation based on the overlap integral with the directly measured bath
spectrum proves the validity of this framework at the weak-coupling
limit. As was already shown numerically in
\cite{our_process_tomography}, the CPMG sequence is found to be
superior to the UDD and CDD sequences, for our collisional bath
spectrum.

As noted above, the weak-coupling assumption is essential for the
overlap-integral formalism to be correct. Under certain experimental
conditions and at low frequencies, our bath does not fulfil this
assumption, and indeed we observe deviations from the corresponding
theoretical calculations. Measuring the coupling spectrum under these
conditions is a major challenge. One way of addressing it is by using
a sequence that mostly resides in the weak-coupling part of the
spectrum, while a small sideband probes the low frequencies of the
spectrum. The idea is to keep most of the overlap between the spectrum
and the filter function in the weak-coupling domain, while still
probing the strong coupling domain of the spectrum. An example of such
a pulse is:
\begin{equation}
\Omega(t)=2\pi f_0 (1+\beta \frac{f_{m}}{f_0} cos(2\pi
f_{m}t)),
\end{equation}
where $f_0$ is the carrier Rabi frequency, $\beta$ is the modulation
index and $f_m$ is the AM modulation frequency. By measuring the
decoherence twice: once with the sideband and once without, it is
possible to extract the coupling spectrum at the sideband frequency,
even if this frequency is in the strong-coupling part of the spectrum.

For a given bath spectrum and a set of constraints on the control
field, an optimal decoupling sequence can be constructed
\cite{gordon:010403,PhysRevLett.104.040401}. This is done by solving the Euler-Lagrange equation which
minimizes the overlap of the control-field filter function and the
bath-coupling spectrum. This procedure is yet to be demonstrated
experimentally. In our system, as well as in other systems, the
spectrum is non-monotonic and as such we expect a non-trivial optimal
decoupling sequence. What can complicate this picture is noise in the
control field. In this work we have circumvented this issue by using
an envelope spectroscopy method. It is possible to extend the
overlap-integral formalism to include the classical noisy control as a
second (classical) bath-like spectral function. The decay rate is then
given by two overlap integrals which can be solved to find the optimal
DD sequence for a noisy control.  Since noise in the control is always
present, this extension of the theory is both necessary and practical.

We acknowledge the financial support of MIDAS, MINERVA, GIF, ISF, and DIP.

\bibliographystyle{nature}

\begin{thebibliography}{10}
\bibitem{remark_collisions}
Collisions are simulated by assuming a steady state density profile and
  calculating each atom's local probability to undergo a collision. Once a
  collision occurred, we need to calculate the velocity of the atoms resulting
  from it. To that end we assume a virtual counterpart only for the purposes of
  this specific collision. The velocity of the virtual counterpart is drawn
  from a Maxwell-Boltzmann distribution modified relative to the velocity of
  the main atom (the one which is really part of the simulation). The collision
  itself is described in these low temperature regime by an s-wave scattering
  process with scattering length of a=$5.2$nm for $^{87}$Rb. We disregard the
  virtual counterpart after the collision. Of course, this process has to be
  proved self-consistent such that it reproduces the Boltzman distribution
  after many such collisions.
\end{thebibliography}

\end{document}